\documentclass{article}
\usepackage[twocolumn]{emulateapj5}

\def\gs{\mathrel{\raise0.35ex\hbox{$\scriptstyle >$}\kern-0.6em 
\lower0.40ex\hbox{{$\scriptstyle \sim$}}}}
\def\ls{\mathrel{\raise0.35ex\hbox{$\scriptstyle <$}\kern-0.6em 
\lower0.40ex\hbox{{$\scriptstyle \sim$}}}}
\eqsecnum

\lefthead{Boughn, Crittenden \& Koehrsen}
\righthead{Large-Scale Structure of the X-ray Background}

\begin{document}
\title{The Large-Scale Structure of the X-ray Background and its Cosmological 
Implications}
\author{S.P. Boughn}
\affil{Department of Astronomy, Haverford College, Haverford, PA  19041
sboughn@haverford.edu}
\author{R.G. Crittenden}
\affil{DAMTP, University of Cambridge, Cambridge CB3 9EW, UK
r.g.crittenden@damtp.cam.ac.uk}
\author{G.P. Koehrsen}
\affil{Department of Astronomy, Haverford College, Haverford, PA 19041
gregory.koehrsen@westtown.edu}

\begin{abstract}
A careful analysis of the $HEAO1~A2~2-10~keV$ full-sky map of the X-ray 
background (XRB) reveals clustering on the scale of several degrees.  After 
removing the contribution due to beam smearing, the intrinsic clustering of the
background is found to be consistent with an auto-correlation function of 
the form
$3.6 \pm 0.9 \times 10^{-4} \theta^{-1}$ where $\theta$ is measured in degrees.
If current AGN models of the hard XRB are reasonable and 
the cosmological constant-cold dark
matter ($\Lambda CDM$) cosmology is correct, 
this clustering implies an X-ray bias factor 
of $b_X \sim 2$.  
Combined with the absence of a correlation between the XRB and
the cosmic microwave background (CMB), this clustering can be used to
limit the presence of an integrated Sachs-Wolfe (ISW) effect and thereby
to constrain the value of the cosmological constant, 
$\Omega_\Lambda \le 0.60$ (95\% C.L.).  This constraint is inconsistent
with much of the $\Omega_\Lambda$ parameter space 
currently favored by other observations.  
Finally, we marginally detect the dipole moment of the 
diffuse XRB and find it to be consistent with the dipole due to
our motion with respect to the mean rest frame of the XRB.  
The limit on the 
amplitude of any intrinsic dipole is 
$\delta I_x / I \le 5 \times 10^{-3}$ at the 95 \% C.L.
When compared to the local bulk velocity, this limit implies a  
constraint on the matter density of the universe of 
${\Omega_m}^{0.6}/b_X(0) \gs 0.24$.
\end{abstract}


\section{Introduction}

The X-ray background (XRB) was discovered before the cosmic microwave
background (CMB), but only now is its origin being fully understood.  
The hard ($2-10 ~ keV$) XRB has been nearly completely resolved into 
individual sources;  most of these are 
active galactic nuclei (AGN), but there is a minor contribution
from the hot, intergalactic medium in rich clusters of galaxies
(Rosati et al. 2002; Cowie et al. 2002; \& Mushotzky et al. 2000).
In addition, the spectra of these faint X-ray sources are consistent with that
of the ``diffuse'' XRB.  If current models of the luminosity functions and
evolution of these sources are reasonably correct, then the XRB arises
from sources in the redshift range $0 < z < 4$,  making them an important
probe of density fluctuations intermediate between relatively nearby 
galaxy surveys ($z \ls 0.5$) and the CMB ($z \sim 1000$).
	
While there have been several attempts to measure large scale,
correlated fluctuations 
in the hard XRB, these have only yielded upper limits or, at best, marginal 
detections (e.g. Barcons et al. 2000, Treyer et al. 1998 and references
cited therein).  
On small scales, a recent correlation analysis of
159 sources in the Chandra Deep Field South survey detected significant 
correlations for separations out to $100~arcsec$ 
(Giacconi et al. 2001).
(At the survey flux level, these sources
comprise roughly two thirds 
of the hard XRB.)  
On much larger scales, a recent analysis by Scharf et al. 
(2000) claims a significant detection of large-scale
harmonic structure in the XRB with spherical harmonic order $1 \le \ell \le 10$ 
corresponding to structures on angular scales of $ \theta \gs 10^\circ$
The auto-correlation results we describe here
complement this analysis,  
indicating clustering on angular scales of $3^{\circ}$ to
$10^{\circ}$, corresponding to harmonic order of $ \ell \ls 30$.
However, all three detections 
have relatively low
signal to noise and require independent confirmation.

The dipole moment of the XRB has received particular attention, primarily 
because of its relation to the dipole in the CMB, which is likely 
due to the Earth's motion with respect to the rest
frame of the CMB.  If this is the case, one expects a similar dipole in the XRB
with an amplitude that is 3.4 times larger because of the difference in
spectral indices of the two backgrounds (Boldt 1987).  In the X-ray literature,
this dipole is widely known as the Compton-Getting effect
(Compton \& Getting 1935). In addition, 
it is quite likely that the XRB has an intrinsic dipole due to the asymmetric
distribution in the local matter density that is responsible for the Earth's 
peculiar motion in the first place.  Searches for both these dipoles have 
concentrated on the hard XRB, since at lower energies the X-ray 
sky is dominated by Galactic structure.  There have been several tentative 
detections of the X-ray dipole (e.g. Scharf et al. (2000)), but these have 
large uncertainties.  
A firm detection 
of an intrinsic dipole or even an upper limit on its presence would 
provide an important constraint on the inhomogeneity of the local distribution 
of matter via a less often used tracer of mass and a concomitant constraint 
on cosmological models (e.g., Lahav, Piran \& Treyer 1997).

This paper is organized as follows.  In section \S2 we describe the 
hard X-ray map used in the analysis, the determination of its 
effective beam size, and cuts made to remove the foreground contaminants. 
In section \S3, we describe the remaining large scale structures 
in the map and the determination of their amplitudes.  The dipole 
is of particular interest, and is the topic of section \S4. 
The correlation function of the residual map and its implications for 
intrinsic correlations are discussed in section \S5.  In section \S6, 
we compare our results to previous observations and discuss the 
cosmological implications of these results in \S7.

\section{HEAO1 A2  ${\it 2 - 10~keV}$ X-ray Map}

There has been much recent progress in understanding the X-ray 
background through instruments such as ROSAT, Chandra and XMM.  
However, these either have too low an energy threshold 
or have too small a field of view to study the large scale structure 
of the hard X-rays.  The best observations relevant to large scale
structure  are still those 
from the HEAO1 A2 experiment that measured the surface brightness of the 
X-ray background in the $0.1 - 60~keV$ band (Boldt 1987).  

\begin{figure}
\centerline{\includegraphics*[height=5.0in,angle=270]{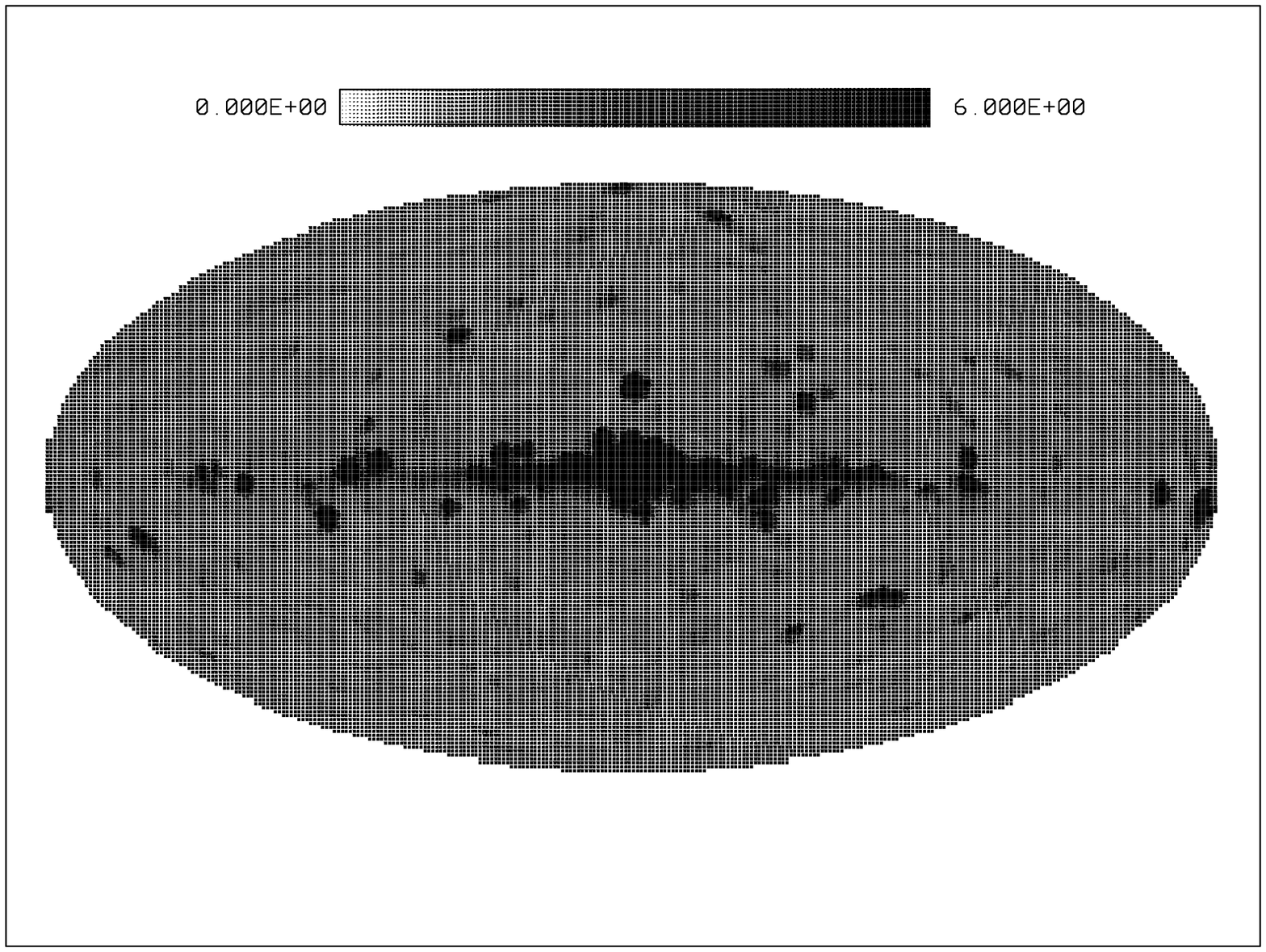}}
\caption{The combined map from the HEAO1 A2 medium and high 
energy detectors, pixelized using the standard COBE quad cubed 
format ($1.3^\circ  \times 1.3^\circ$ pixels.) The effective beam 
size is approximately $3^\circ$. The most visible features, the  
Galactic plane and the nearby bright sources, are removed 
from the maps we analyze. } 
\label{fig:heao} 
\end{figure}

The HEAO1 data set we 
consider was
constructed from the output of two medium energy detectors (MED) with different
fields of view ($3^\circ \times 3^\circ$ and $3^\circ \times 1.5^\circ$) and 
two high energy detectors (HED3) with
these same fields of view.  
These data were collected during the six month period
beginning on day 322 of 1977.  Counts from the four detectors were combined and 
binned in 24,576  $1.3^\circ \times 1.3^\circ$ pixels. 
The pixelization we use is an equatorial 
quadrilateralized spherical cube projection on the sky, 
the same as used for the COBE satellite CMB maps 
(White and Stemwedel 1992).
The combined map has a spectral bandpass (quantum efficiency $\gs 50\%$) of 
approximately $3-17~keV$ (Jahoda \& Mushotzky 1989) and is shown in Galactic 
coordinates in Figure \ref{fig:heao}. For consistency with other work, all 
signals are converted to equivalent flux in the $2-10~keV$ band.

Because of the ecliptic longitude scan pattern of the HEAO satellite, sky
coverage and therefore photon shot noise are not uniform.  However, the
variance of the cleaned, corrected map, 
$2.1 \times 10^{-2}~(TOT~counts~s^{-1})^2$, 
is much larger than the variance of photon shot noise, 
$0.8 \times 10^{-2}~(TOT~counts~s^{-1})^2$, where 
$1~TOT~counts~s^{-1} \approx 2.1 \times 10^{-11} 
erg~s^{-1} cm^{-2}$ (Allen, Jahoda \& Whitlock 1994).  
This implies that most of the variance in
the X-ray map is due to ``real'' structure.  For this reason and 
to reduce contamination from 
any systematics that might be correlated with the scan 
pattern, we chose to weight the pixels equally in this analysis.

\subsection{The point spread function} 

\begin{figure}
\centerline{\includegraphics*[height=3.0in,angle=270.]{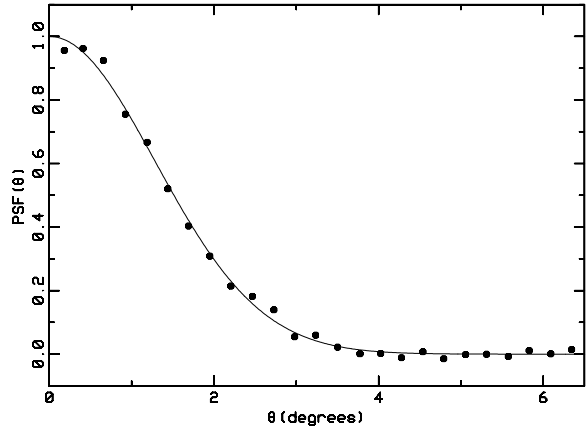}}
\caption{The mean point spread function for the 
combined map found by averaging the individual PSFs of sixty 
strong HEAO1 point sources.  The data is well fit 
by a Gaussian with FWHM of $3.04^\circ$.} 
\label{fig:psf} 
\end{figure}

To determine the level of intrinsic correlations, 
we must account for the effects of
beam smearing and so it is essential to characterize the point spread
function (PSF) of the above map.  The PSF varies somewhat with 
position on the sky because of the pixelization and the   
asymmetric beam combined with the HEAO1 scan pattern.  We obtained 
a mean PSF by averaging the individual PSFs of 60 strong HEAO1 point sources 
(Piccinotti 1982) that were located
more than $20^{\circ}$ from the Galactic plane.  
The latter condition was imposed 
to avoid crowding and to approximate the windowing of the subsequent 
analysis (see \S2.2).  The composite PSF, shown in Figure \ref{fig:psf},
is well fit by a Gaussian with a full 
width, half maximum (FWHM) of $3.04^\circ$.

As a check of this PSF, we generated
Monte Carlo maps of sources 
observed with $3^\circ \times 3^\circ$ and $3^\circ \times 1.5^\circ$ (FWHM)
triangular beams appropriate for the A2 detectors (Shafer 1983) 
and then combined the maps with quadcubed pixelization as above.
The resulting average PSF from these trials is also well fit by a Gaussian with
a FWHM of $2.91^\circ$, i.e., about $4.5\%$ less than that in the Figure 
\ref{fig:psf}.  
Considering that the widths of the triangular beams given above are nominal, 
that the triangular beam pattern is only approximate (especially at higher 
energies), and that we did not take into account the slight
smearing in the satellite scan 
direction (Shafer 1983), the agreement is remarkably good.  
In the following analysis, we use the $3.04^\circ$ fit derived 
from the observed map; however,
changing the PSF FWHM by a few percent does not significantly
affect the results of this paper.

\subsection{Cleaning the map} 

To remove the effects of the Galaxy and strong 
extra-galactic point sources, some regions of the map were 
excluded from the analysis.  The dominant feature in the HEAO map is the 
Galaxy (see Figure \ref{fig:heao}) so all data within 
$20^\circ$ of the Galactic plane or within $30^\circ$ of the 
Galactic center were cut from the map.  In addition, 
large regions ($6.5^\circ \times 6.5^\circ$) centered
on $92$ discrete X-ray sources with $2 - 10~keV$ fluxes larger than 
$3 \times 10^{-11} erg~s^{-1} cm^{-2}$ (Piccinotti 1982) were removed from the 
maps.  Around the sixteen brightest of these sources (with fluxes larger than
$1 \times 10^{-10} erg~s^{-1} cm^{-2}$) the cut regions were enlarged to 
$9^\circ \times 9^\circ$.  Further enlarging 
the area of the excised regions had a negligible effect on the following 
analysis so we conclude that the sources have been effectively removed.
The resulting ``cleaned'' map (designated Map A) has a sky coverage of 
$55.5\%$ and is our baseline map for further cuts.  

To test the possibility of further point source 
contamination, we also used the
ROSAT All-Sky Survey (RASS) Bright Source Catalog (Voges et al. 1996) 
to identify relatively bright sources.
While the RASS survey has somewhat
less than full sky coverage ($92\%$), it has a relatively low flux limit that 
corresponds to a $2-10~keV$ flux of 
$\sim 2 \times 10^{-13}~erg~s^{-1}~cm^{-2}$ 
for a photon spectral index of $\alpha = -2$.  
Every source in the RASS catalog was
assigned a $2-10~keV$ flux from its B-band flux by assuming a spectral 
index of $-3< \alpha < -1$ as deduced from its HR2 hardness
ratio.  For fainter sources, the computed
value of $\alpha$ is quite uncertain; if it fell outside the typical range 
of most X-ray sources, $-3< \alpha < -1$, 
then $\alpha$ was simply forced to be $-1$ or $-3$.  
It is clear that extrapolating RASS flux
to the $2-10~keV$ band is not accurate, so one must consider the level to
which sources are masked with due caution.  However, we are only using these
fluxes to mask bright sources and so this procedure is unlikely to bias the 
results.  

We considered maps where the ROSAT sources were removed at three different 
inferred $2-10~keV$ flux thresholds. 
First, we identified sources with fluxes exceeding  
the Piccinotti level, $3 \times 10^{-11} erg~s^{-1} cm^{-2}$. 
Thirty-four additional, high Galactic latitude RASS sources were removed, 
resulting in a map with sky coverage of $52\%$
(designated Map B).
In order 
to compare more directly with the results of Scharf et al. (2000) (see \S6)
we removed sources at their flux level, 
$2 \times 10^{-11} erg~s^{-1} cm^{-2}$.  
The map masked 
in this way has $47\%$ sky coverage (compared to the $48\%$ coverage of 
the Scharf et al. analysis) and is designated Map C. 
Finally, to check how sensitive our dipole results are to 
the particular masking
of the map, we lowered the flux cut level to 
$1 \times 10^{-11} erg~s^{-1} cm^{-2}$,
which reduced the sky coverage to $34\%$.  The map resulting from this cut
is designated Map D in Table 3.

As an alternative to using the RASS sources, 
the map itself was searched for ``sources'' that 
exceeded the nearby background by a specified amount.  
Since the quad-cubed format lays out the pixels on an approximately square
array, we 
averaged each pixel with its eight neighbors 
and then compared this value with the median value of the next nearest
sixteen pixels (ignoring pixels within the masked regions).  If the
average flux associated with a given pixel exceeded the median flux of the
background by a prescribed threshold, 
then all 25 pixels ($6.5^\circ \times 6.5^\circ$) were 
removed from further consideration.  
For a threshold 
corresponding to 2.2 times the mean shot noise in the map approximately 
120 more ``sources'' were identified and masked resulting in a sky coverage of
$42\%$.  This map is labeled Map E in Table 3.
Finally, we used an even more aggressive cut corresponding to 1.75 times the
mean shot noise which resulted in a masked map with $33\%$ sky coverage.
This map is labeled Map F.

\section{Modeling the Local Large-Scale Structure }

\subsection{Sources of large scale structure} 

There are several local sources of large-scale structure in the HEAO map 
which can not be eliminated by masking isolated regions.
These include diffuse emission from the Galaxy, emission (diffuse and/or 
faint point sources) from the Local Supercluster, the Compton-Getting dipole, 
and a linear time drift in detector sensitivity.  Since none of these 
are known \textit{a priori}, we fit an eight parameter model to the data.   
Of course, the Compton-Getting dipole is known in 
principle if one assumes the kinetic origin 
of the dipole in the cosmic microwave background; however, there may also be an 
intrinsic X-ray dipole that is not accounted for. (See \S4 below.)   
Only one correction was made \textit{a priori} to the map and that was for the 
dipole due to the Earth's motion around the sun; however, 
this correction has a negligible effect on the results.  
A more detailed account of the model is given in 
Boughn (1999).

The X-ray background has a diffuse (or unresolved) Galactic component which 
varies strongly with Galactic latitude (Iwan et al. 1982).  This emission is
still significant at high Galactic latitude ($b_{II}>20^\circ$) and extrapolates
to $\sim 1\%$ at the Galactic poles.  We modeled this emission in two ways.  The
first model consisted of a linear combination of a secant law Galaxy with 
the Haslam $408~GHz$ full sky map (Haslam et al. 1982). 
The latter was included to take into 
account X-rays generated by inverse Compton scattering of CMB photons 
from high energy electrons in the Galactic halo, the source of much of 
the synchrotron emission in the Haslam map.  As an alternative 
Galaxy model we also considered the two disk, exponentially truncated model
of Iwan et al. (1982).  Our results are independent of which model is used.

In addition to the Galactic component, evidence has been found for faint 
X-ray emission from plane of the Local Supercluster  
(Jahoda 1993, Boughn 1999). 
Because of its faintness, very detailed models
of this emission are not particularly useful.  The model we use here is a simple
``pillbox'', i.e uniform X-ray emissivity within a circular disk of thickness 
equal to a $1/4$ of the radius and with its center located 
$4/5$ of a radius from us in
the direction of the Virgo cluster (see Boughn 1999 for details).  The amplitude
of this emission, while significant, is largely independent of the details
of the model and, in any case, has only a small effect on the results. 

Time drifts in the detector sensitivity can also 
lead to apparent structure in 
the reconstructed X-ray map. 
At least one of the A-2 detectors changed sensitivity by $\sim 1\%$ in the six 
month interval of the current data set (Jahoda 1993).  Because of the ecliptic
scan pattern of the HEAO satellite, this results in a large-scale pattern
in the sky which varies with ecliptic longitude with a period of $180^\circ$.
If the drift is assumed to be linear, the form of the resulting large-scale 
structure in the map is completely determined.  A linear drift of unknown 
amplitude is taken into account by constructing a sky map with the appropriate 
structure and then fitting for the amplitude simultaneously with the other 
parameters.  We investigated the possibility of non-linear drift by considering
quadratic and cubic terms as well; however, this did not significantly reduce
the $\chi^2$ of the fit nor change the subsequent results.

\subsection{Modeling the maps} 

The eight parameters that characterize the amplitude of these 
structures are used
to model the large-scale structure in the HEAO map.
Let the X-ray
intensity map be denoted by the vector $\bf{I}$, where the element $I_i$ 
is the intensity in the $i^{th}$ pixel.
The observed intensity is modelled as the sum of eight templates
with amplitudes described by the eight dimensional vector $\bf{a}$, 
\begin{equation}
{\bf{I}} = \tilde{X}{\bf{a}} + {\bf{n}} 
\end{equation} 
where $\tilde{X}$ is an $8 \times n_{pix}$ matrix whose elements are the 
values of each template function at each pixel of the map.  
As discussed above, these template functions include: a uniform map to 
represent the monopole of the X-ray background; 
the three components of a 
dipole (in equatorial coordinates); 
the large-scale pattern resulting from a linear 
instrumental gain drift;  a Galactic secant law; the Haslam 
$408~GHz$ map ; and the amplitude of
the ``pillbox'' model of the local supercluster.  
The noise vector ${\bf{n}}$ is assumed to be Gaussian distributed with 
correlations described by $\tilde{C}\equiv \langle {\bf{n \, n}}^T\rangle$.  

As discussed above (\S2), we chose to weight each pixel equally  
since the shot noise is considerably less than the ``real'' fluctuations
in the sky.  
For the purposes of fitting the map to the 
above model we consider both photon shot noise and fluctuations in the XRB
(see Figure \ref{fig:acf})
to be ``noise''.  This noise is correlated and a minimum
$\chi^2$ fit must take such correlations into account.  
However, for simplicity, we ignore these correlations when finding the best 
fit model amplitudes and perform a  
standard least squares fit by minimizing $|{\bf{I}} - \tilde{X}{\bf{a}}|^2$
on the cleaned HEAO map. 
From the standard equations of linear regression the values of the parameters 
that minimize this sum are 
\begin{equation}
{\bf{a}} = \tilde{B}^{-1} \tilde{X}^T {\bf{I}}
\end{equation}
where $\tilde{B} = \tilde{X}^T \tilde{X}$ is a symmetric 
eight by eight matrix.
This would be the maximum likelihood estimator if the correlation matrix 
were uniform and diagonal.  
Though this fit ignores correlations in the errors, it is unbiased and 
is likely to be very close to the minimum $\chi^2$ (maximum likelihood) 
fit, since 
the noise correlations are on a much smaller scale than the features we 
are attempting to fit. 

The correlated nature of the noise cannot be ignored when computing 
the uncertainties in the fit since there are far fewer noise independent data 
points than there are pixels in the map.  It is straightforward to show that 
errors in the estimated parameters $\delta {\bf{a}}$ are given by 
\begin{equation}
\label{eqn:sigc}
\langle \delta {\bf{a}} \, \delta {\bf{a}}^T \rangle = 
\tilde{B}^{-1} \tilde{X}^T \tilde{C} \tilde{X} \tilde{B}^{-1} 
\end{equation}
This error is likely to be only slightly larger than 
would be the case for the maximum likelihood estimator.  
$\tilde{C}$
is a combination of the uncorrelated shot noise and the correlated 
fluctuations indicated in Figure \ref{fig:acf}.
We assume it to be homogeneous and
isotropic, i.e., that $\tilde{C}_{ij}$ depends only on the angular 
separation of the 
$i$ and $j$ pixels.  

\begin{table*}[ht]
\begin{center}
\begin{tabular}{l|l|c|c}
\multicolumn{1}{l}{}
& \multicolumn{1}{l}{Parameter} & \multicolumn{1}{l}{Uncorrected} &
\multicolumn{1}{l}{Corrected for C-G} \\
\cline{1-4}
$a_1$ & background & 328.6 $\pm$ 1.9 & same \\
$a_2$ & $\hat{x}$ dipole & -1.17 $\pm$ 0.62 & -0.24 $\pm$ 0.62 \\
$a_3$ & $\hat{y}$ dipole & -0.38 $\pm$ 0.98 & -0.68 $\pm$ 0.98 \\
$a_4$ & $\hat{z}$ dipole & -0.52 $\pm$ 0.69 & -0.34 $\pm$ 0.69 \\
$a_5$ & time drift & 7.15 $\pm$ 1.23 & same \\
$a_6$ & secant law & 3.28 $\pm$ 0.84 & same \\
$a_7$ & Haslam map & 0.03 $\pm$ 0.08 & same \\
$a_8$ & Supercluster & 4.11 $\pm$ 1.35 & same
\end{tabular}
\end{center}
\caption{}{Eight fit parameters for Map C (sources brighter than 
$2\times10^{-11}~erg~s^{-1}cm^{-2}$ removed).  The units are $0.01~
TOT~count~s^{-1} (4.5~deg^2)^{-1} \simeq 1.54\times 10^{-10}erg~s^{-1}cm^{-2}$.
Fits are shown both for the original map and for the map corrected for the
Compton-Getting (C-G) dipole.  
}
\label{tab:comp}
\end{table*}

Table \ref{tab:comp}
lists the values and errors of the 
parameters fit to Map C (see \S2).  Instrument time drift, the Galaxy 
and structure associated with the local supercluster
all appear to be significant detections.  The dipole 
is detected at about the $2~\sigma$ level and is consistent
with that expected for the Compton-Getting dipole (see Table \ref{tab:dipole}).
Table \ref{tab:corrm}
lists the elements of the normalized correlation matrix of the 
fit parameters and 
it is apparent that the parameters are largely uncorrelated.  This was 
supported by fits that excluded some of the parameters (see \S4).  

\begin{table*}[ht]
\begin{center}
\begin{tabular}{c|rrrrrrrr}
\multicolumn{1}{c}{ } & \multicolumn{1}{c}{$a_1$} & \multicolumn{1}{c}{$a_2$} &
\multicolumn{1}{c}{$a_3$} & \multicolumn{1}{c}{$a_4$} &
\multicolumn{1}{c}{$a_5$} & \multicolumn{1}{c}{$a_6$} & 
\multicolumn{1}{c}{$a_7$} & \multicolumn{1}{c}{$a_8$} \\ \cline{2-9}
$a_1$ &  1.0 &	0.0 & -0.5 &  0.1 & -0.4 & -0.6 & -0.1 & -0.3 \\
$a_2$ &  0.0 &  1.0 & -0.1 &  0.2 &  0.3 & -0.1 &  0.0 &  0.1 \\
$a_3$ & -0.5 & -0.1 &  1.0 &  0.1 &  0.0 &  0.7 &  0.0 & -0.3 \\
$a_4$ &  0.1 &  0.2 &  0.1 &  1.0 & -0.1 &  0.0 &  0.0 & -0.1 \\
$a_5$ & -0.4 &  0.3 &  0.0 & -0.1 &  1.0 &  0.0 &  0.0 &  0.2 \\
$a_6$ & -0.6 & -0.1 &  0.7 &  0.0 &  0.0 &  1.0 &  0.0 & -0.4 \\
$a_7$ & -0.1 &  0.0 &  0.0 &  0.0 &  0.0 &  0.0 &  1.0 & -0.3\\
$a_8$ & -0.3 &  0.1 & -0.3 & -0.1 &  0.2 & -0.4 & -0.3 &  1.0
\end{tabular}
\end{center}
\caption{}
\label{tab:corrm}
\begin{center}
{Normalized correlation coefficients for the fit parameters in Table \ref{tab:comp}.}
\end{center}
\end{table*}

To compute the true 
$\chi^2 \equiv ({\bf{I}} - \tilde{X}{\bf{a}})^T \tilde{C}^{-1} 
({\bf{I}} - \tilde{X}{\bf{a}})$ 
of the fit requires inverting $\tilde{C}$ 
which is a $11,531 \times 11,531$ matrix.  Instead we compute an effective
reduced $\chi^2$ using
\begin{equation}
\chi_{eff}^2 \equiv {1 \over N} ({\bf{I}} - \tilde{X}{\bf{a}})^T \tilde{D}^{-1}
({\bf{I}} - \tilde{X}{\bf{a}})
\end{equation}
where $\tilde{D}$
is the diagonal part of the 
correlation matrix, $\tilde{D}_{ii} = \sigma_{s,i}^2 +\sigma_b^2$,  
$\sigma_{s,i}$ is the shot noise in the $i^{th}$ pixel,
${\sigma_b}^2$ is the variance of the fluctuations in the XRB, and $N$ is the 
number of pixels minus eight, the number of degrees of 
freedom in the fit.  The shot noise in a given pixel is 
inversely proportional to the number of photons received and we assume
is inversely proportional to the coverage of that pixel.  This is approximately 
true since all the non-flagged pixels are exposed to approximately the same 
flux.  We find ${\chi_{eff}}^2 = 1.00$ for this fit, 
which we take as an indication that 
we have properly characterized the amplitude of the noise so that the errors 
quoted in the table have neither been underestimated nor overestimated.
However, it should be emphasized that ${\chi_{eff}}^2$ is not to be interpreted
statistically as being derived from a ${\chi}^2$ distribution.

The residual maps show 
very little evidence 
for structure on angular scales $\theta > 10^{\circ}$ above the level of
the noise,   
$\langle \delta I^2 \rangle / \bar{I}^2 \sim 10^{-5}$ where $\delta I$
are the residual fluctuations in X-ray intensity and $\bar{I}$ is the mean 
intensity (see Figure \ref{fig:resid}).  Since all the components of the model 
have significant structure on large angular scales, it appears that these
particular systematics have been effectively eliminated.

\section{The Dipole of the X-ray Background}

The dipole fit to the map is consistent with the Compton-Getting
dipole and there is no evidence for any additional intrinsic dipole in the
XRB.  To make this more quantitative, we corrected the maps for the
predicted Compton-Getting dipole and fit the corrected 
map for any residual, intrinsic dipole.  These dipole fit parameters are 
also included in Table \ref{tab:comp}. 

Leaving out any individual model component, such as 
the time drift, the galaxy, Haslam or supercluster template, 
made little difference 
in the amplitude of the fit dipole.  This is, perhaps, not too surprising
since the Galaxy and time drift models are primarily quadrupolar in nature
and the pancake model, while possessing a significant dipole moment, has 
a relatively small amplitude.  All such fits were consistent with the 
Compton-Getting dipole alone.  Even when all four of these parameters were
excluded from the fit, the dipole amplitude increased by only 
$0.004~TOT~counts~s^{-1}$ with a direction that was was $33^{\circ}$ from
that of the CMB dipole.  The effective $\chi^2$ for the four parameter 
fit was, however, significantly worse, i.e., ${\chi_{eff}}^2 = 1.05$.

Table \ref{tab:dipole} lists the amplitude and direction of the dipole fit
to Map C along with the fits to Maps D, E, and F.  All of these fits are
consistent with amplitude and direction of the Compton-Getting dipole (as
inferred from the CMB dipole) which is also indicated in the Table.
The effective $\chi^2$s of these fits range from 0.99 to 1.01, 
again indicating
that the amplitude of the noise is reasonably well characterized.  No
errors are given for these quantities for reasons that will be discussed below.
In order to check for unknown systematics, we performed dipole fits to a variety
of other masked maps with larger Galaxy cuts as well as cuts
of the brighter galaxies in the Tully Nearby Bright Galaxy Atlas (Tully 1988).
The details of these cuts are discussed in Boughn (1999); however, none had
a significantly different dipole fit.

Since all of these dipoles are consistent with the Compton-Getting dipole we
also fit these maps with a six parameter fit in which the dipole direction was
constrained to be the direction of the CMB dipole.  The dipole amplitude of
these fits and errors computed according to Eq. (3-3)
are also given in Table \ref{tab:dipole}.  

\begin{table*}[ht]
\begin{center}
\begin{tabular}{l|c|c|c||c}
\multicolumn{1}{l}{Map} & \multicolumn{1}{l}{Dipole } 
& \multicolumn{1}{l}{$l_{II}$} & \multicolumn{1}{l}{$b_{II}$}
& \multicolumn{1}{l}{Constrained Dipole }\\
\cline{1-5}
Map C   &     0.0133  &  $309^{\circ}$ &  $39^{\circ}$ & 0.0117 $\pm$ 0.0064 \\
Map D   &     0.0218  &  $300^{\circ}$ &  $33^{\circ}$ & 0.0184 $\pm$ 0.0062 \\
Map E   &     0.0150  &  $296^{\circ}$ &  $50^{\circ}$ & 0.0148 $\pm$ 0.0059 \\
Map F   &     0.0190  &  $283^{\circ}$ &  $44^{\circ}$ & 0.0184 $\pm$ 0.0064 \\
C-G     &     0.0145  &  $264^{\circ}$ &  $48^{\circ}$ & 0.0145

\end{tabular}
\end{center}
\caption{}
\label{tab:dipole} 
{
The dipole amplitude and directions are from the 8-parameter fits  in
$TOT$ units and Galactic coordinates. Map C is for the map of
Table \ref{tab:comp}; 
Map D is masked at a source level of $\sim1 \times 10^{-11}
erg~s^{-1} cm^{-2}$; Map E is masked with internal source identification;
and Map F is masked with a lower level of internal source 
identification (see \S2 for full details).  Also listed are the amplitude
and direction of the Compton-Getting dipole (G-P) as inferred from
the CMB dipole.
The constrained amplitudes are for dipole models fixed
to the direction of the CMB dipole. 
}
\end{table*}

Even though we find no evidence for an intrinsic dipole in the XRB, it 
would be useful to place an upper limit on its amplitude. 
We define the dimensionless dipole by writing the first two moments of 
the X-ray intensity as 
\begin{equation} 
I(\hat{\bf n}) = \bar{I} (1 + \vec{\Delta}  \cdot \hat{\bf n}),  
\end{equation} 
where $\vec{\Delta}$ is a vector in the direction of the dipole. 
There are various approaches one could take to find an upper limit,
and the problem 
is complicated somewhat because the error bars are anisotropic (see Table 1).  
The dipole in the $\hat{y}$ direction is less constrained than in the 
other directions because of the anisotropic masking of the map. 
Here we take the limits on the individual components of the intrinsic dipole
and marginalize over the dipole direction to obtain a distribution for its 
amplitude.  For this, we use a Bayesian formalism and assume a uniform 
prior on the amplitude, $|\vec{\Delta}|$.   
We find $\Delta < 0.0052$ at the 
95 \% C.L. If the direction of the dipole is fixed to be that of the CMB
dipole then the 95\% C.L. upper limits on the dipole amplitudes fall in the
range $0.0030$ to $0.0043$ for the fits listed in
Table \ref{tab:dipole}.

The same sort of problem arises when trying to attach an error bar to the 
amplitude of the dipole fits to the maps which include the Compton-Getting
dipole.  It seems clear from Table \ref{tab:comp} that we find 
evidence for a dipole at
the 2 $\sigma$ level.  
This is supported by the six parameter fits of Table \ref{tab:dipole}, 
where the various maps indicate positive detections at a 2 to 3 $\sigma$ level.
However, in the eight parameter fits, the dipole amplitude is a non-linear 
combination of the fit components.  
There are two approaches we can take in converting the three dimensional 
limits to a limit on the dipole amplitude.  We can either fix the direction 
in the direction of the CMB dipole, which results in the constrained limits
shown in Table \ref{tab:dipole}.  Alternatively, we can marginalize over 
the possible directions of the dipole, which will necessarily result 
in weaker limits than when the direction is fixed.  
This is particularly true here, where the direction of greatest uncertainty 
in the dipole measurement is roughly orthogonal to the expected dipole 
direction.  
In the case of the Compton-Getting dipole, there is a strong 
prior that it should be in the CMB dipole 
direction, so our limit is stronger than
it would be if we did not have information about the CMB. 

In addition to the upper limit on the intrinsic dipole amplitude,
we can also constrain the underlying dipole variance, which
can, in turn, be used to test theoretically predicted power spectra.
While the observed amplitude is related to the dipole variance, 
$\langle \Delta^2 \rangle = 3 \sigma_\Delta^2$,
there is large uncertainty due to cosmic variance.  
The dipole represents only three independent samplings of $\sigma_\Delta$.
To constrain $\sigma_\Delta$, we again take a Bayesian approach and calculate
the likelihood of observing the data given the noise and $\sigma_\Delta$,
\begin{equation}
{\cal {P}} (\vec{\Delta}|\sigma_\Delta) \propto \prod
e^{- \Delta_i^2/2(\sigma_i^2 + \sigma_\Delta^2)} 
(\sigma_i^2 + \sigma_\Delta^2)^{-1/2},
\end{equation} 
where the product is over the three spatial directions and 
we have ignored the small off-diagonal
noise correlations (see Table \ref{tab:corrm}).
With a uniform prior on $\sigma_\Delta$, its posterior distribution implies
a 95\% C.L. upper limit of $\sigma_\Delta < 0.0064 $.
This is twice as high as would be inferred from the limit on the dipole because
of the significant tail in the distribution due to cosmic variance.
The limit implied by the dipole (of Map C), $\sigma_\Delta = \Delta/\sqrt{3} < 0.0030 $ is
at the 80\% C.L.  
The difference between the limits
arises because occasionally a small dipole can occur
even when the variance is large. 
  
The bottom line
is that we have detected the dipole in the XRB at about the 2 $\sigma$
level and that it is consistent with the Compton-Getting dipole.  There is
no evidence for any other intrinsic dipole at this same level.  We will
discuss the apparent detection of an intrinsic dipole by Scharf et al. (2000)
in \S6.

\section{Correlations in the X-ray Background}

\begin{figure}
\centerline{\includegraphics*[height=3.0in,angle=270.]{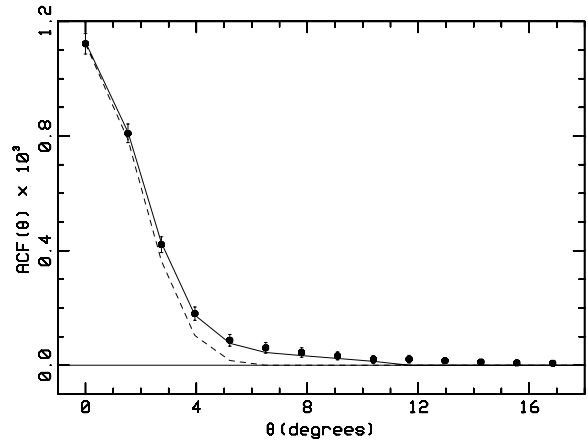}}
\caption{The auto-correlation function of the HEAO1 A2 map
with bright sources and the Galactic plane removed and corrected for
large-scale, high Galactic latitude structure.
The dashed curve is that expected from beam smearing due to the PSF
of the map while the solid curve includes a contribution due to
clustering in the XRB (see \S5).}
\label{fig:acf}
\end{figure}

\begin{figure}
\centerline{\includegraphics*[height=3.0in,angle=270.]{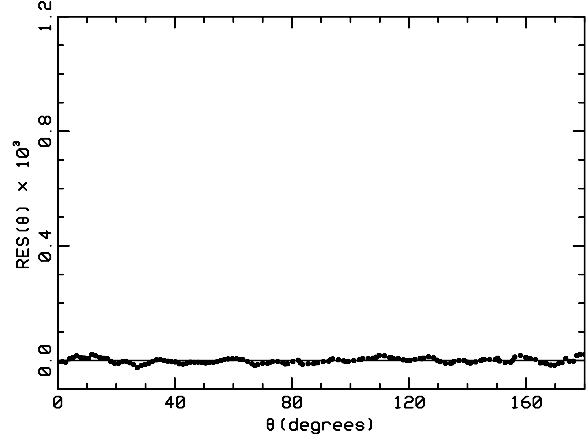}}
\caption{The residuals of the ACF fit from Figure 3, after the shot noise, 
PSF and a simple model of the intrinsic fluctuations have been removed. 
} 
\label{fig:resid} 
\end{figure}

\begin{figure}
\centerline{\includegraphics*[height=3.0in,angle=270.]{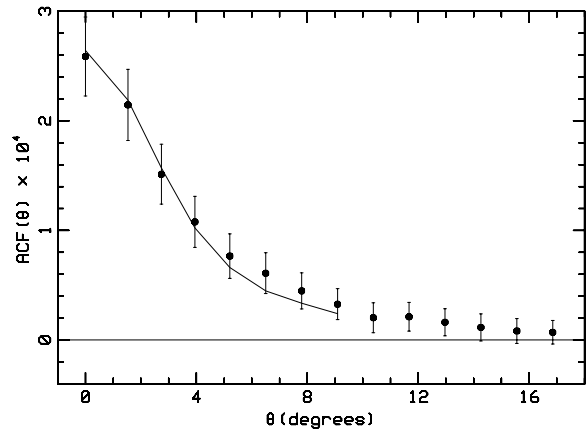}}
\caption{The intrinsic ACF, with shot noise and PSF fits removed. 
For comparison, a simple $\theta^{-1}$ model for the intrinsic 
correlations is shown.  The data beyond $9^\circ$ is not used because of 
uncertainty due to the fitting of the large scale structures. 
The model has been smoothed by the PSF and 
corrected for the removal of the large scale 
structures, which suppresses the correlations on scales larger than 
$10^\circ.$}  
\label{fig:intrin} 
\end{figure}

A standard way to detect the clustering of sources (or
of the emission of these sources) is to compute the 
auto-correlation function (ACF), defined by
\begin{equation}
\omega(\theta) = {1 \over \bar{I}^2} \sum_{i,j} (I_i -\bar{I})
(I_j-\bar{I}) / N_{\theta}
\end{equation}
where the sum is over all pairs of pixels, $i,j$, separated by an angle 
$\theta$, $I_i$ is the intensity of the $ith$ pixel,
$\bar{I}$ is the mean intensity, and $N_{\theta}$ is the number of 
pairs of pixels separated by $\theta$. Figure \ref{fig:acf} shows
the ACF of the residual map after being corrected with the 8-parameter fit 
and for photon shot noise in 
the $\theta = 0^\circ$ bin.  
The error bars are highly correlated and were determined from Monte Carlo trials
in which the pixel intensity distribution was assumed to be Gaussian
with the same ACF as in the figure.
There is essentially no significant structure for
$\theta > 13^{\circ}$ once local structures have been removed, 
as is evident in Figures \ref{fig:acf}, \ref{fig:resid}, and \ref{fig:intrin}.

It is clear from Figure \ref{fig:acf} that the residuals of Map A 
possess significant 
correlated structure. 
It must be determined how much, if any, is due 
to clustering in the XRB and how much is simply due to smearing by the PSF of 
the map. It is straightforward to show that an uncorrelated signal
smeared by a Gaussian PSF,
$PSF(\theta) \propto e^{-\theta ^2 / 2\sigma_p ^2}$, results in an ACF of
the form $\omega (\theta) \propto e^{-\theta ^2 /4\sigma_p ^2}$ where 
$\sigma_p = 1.29^{\circ}$ is the Gaussian width of the PSF in 
Figure \ref{fig:psf}  
($\theta_{FWHM}^2  = 8 \sigma_p^2 \ln 2.$) 
The dashed curve in Figure \ref{fig:acf}
is essentially this functional form, modified
slightly to take into account the pixelization.  
In the plot, its amplitude has 
been forced to agree with $\theta = 0^\circ$ data point, while a maximum likelihood 
fit results in an 
amplitude about 5\% lower  
(a consequence of the correlated noise). 
For $\theta \gs 3^{\circ}$, the ACF of the data clearly exceeds  
that accountable by beam smearing
and this excess is even more pronounced with the maximum likelihood fit.
The reduced $\chi^2$ for the fit to the first
eight data points ($\theta \ls 9^{\circ}$) is  $\chi ^2 = 18.6$ 
for six degrees
of freedom which is another measure of the excess structure between 3 and 9
degrees. Note, this is a two parameter fit since the photon shot noise (which 
occurs only at $\theta = 0^\circ$) is also one of the parameters.

While it is apparent that there is some intrinsic correlation in the X-ray 
background, it cannot be estimated by the residual 
to the above two parameter fit 
since that overestimates the contribution of beam smearing in order
to minimize $\chi^2$.  Instead, we also include in the fit 
a form for the intrinsic
correlation and find its amplitude as well. 
Since the signal to noise is too small to allow a 
detailed model of the intrinsic clustering, 
we chose to model it with a
simple power-law $\omega (\theta) =  (\theta_0/\theta)^\epsilon$.  
This form provides 
an acceptable fit to the ACF of both radio and X-ray surveys on somewhat smaller
angular scales 
(Cress \& Kamionkowski 1998, Soltan et al. 1996, Giacconi et al. 2001).  
This intrinsic correlation was then convolved with the PSF and applied to the 
quadcube pixelization of the map (e.g., Boughn 1998).  

Finally, it is important 
to take into account the effects of the 8-parameter fit used to remove the
large-scale structure as discussed in \S3.  If the X-ray 
background has intrinsic structure on the scale of many degrees, the 
8-parameter fit will tend to remove it in order to minimize $\chi^2$.  Since 
the model is composed of relatively large scale features, the greatest effect 
is expected for the largest angles.  The significance of this effect was 
determined by generating Monte Carlo
trials assuming a Gaussian pixel intensity distribution with the same ACF 
as in Figure \ref{fig:acf}.  
The 8-parameter model was then fit and each trial map was 
corrected accordingly.  The ACFs computed for these corrected maps
indicate that, as expected, the value of the ACF is significantly attenuated 
for larger angles.  The attenuation factor for $\theta = 9^{\circ}$ is
already 0.55 and decreases rapidly for larger angles.  The errors indicated in
Figure \ref{fig:acf} were also determined from these Monte Carlo trials and, as
mentioned above, are highly correlated.  

We model the auto-correlation function as a sum of three templates and fit 
for their best amplitude.  Much of the analysis parallels the discussion for 
the fits of large scale structure in the map, with the exception that here 
the number of bins is small enough that it is simple to calculate the 
maximum likelihood fit.  
Again, we model the observed correlation function vector as 
${\bf{\omega}} = \tilde{W} {\bf{c}} + {\bf{n}}_{\omega}$, 
where $\tilde{W}$ is a $3 \times n_{bin}$ matrix containing the templates 
for shot noise ($\omega_s$), beam smearing ($\omega_{PSF}$), 
and intrinsic correlations in the XRB ($\omega_{intr}$).  
The amplitudes are 
given by the three element vector ${\bf{c}}$ and the noise is described 
by the correlation matrix determined from Monte Carlo trials, 
$\tilde{C}_\omega = \langle{\bf{n}}_{\omega} {\bf{n}}_{\omega}^T \rangle$. 

Shot noise contributes only to the first bin (zero separation) of the ACF
and has amplitude given by its variance, $\omega_s$.  Beam smearing 
contributes to the ACF with a template that looks like the beam convolved with 
itself, so appears as a Gaussian with a FWHM a factor of $\sqrt{2}$ larger 
than that of the beam and has amplitude denoted as $\omega_{psf}$. 
Finally, the intrinsic correlations are 
modelled as $(\theta_0/\theta)^\epsilon$, 
which is then smoothed appropriately by the beam.  Its amplitude is 
denoted by its inferred correlation at zero separation, $\omega_{intr}.$
We fit using a range of indices, $0.8 \ge \epsilon \ge 1.6$, which cover
the range of theoretical models of the intrinsic correlation. 
Both the PSF template and the intrinsic template are modified to include 
the effects of the attenuation at large angles as discussed above.    
 
Minimizing $\chi^2$ with respect to the three fit parameters, 
${\bf{c}}$, results 
in the maximum likelihood fit to the model if one assumes Gaussian statistics.  
This assumption is reasonable by virtue of the central limit theorem 
since each data point consists of the combination of the signals from a 
great many pixels, each of which is approximately Gaussian distributed.
In the presence of correlated noise, $\chi^2$ 
is defined by 
\begin{equation}
\chi^2 =  ({\bf{\omega}} - \tilde{W} {\bf{c}})^T \tilde{C}^{-1}_{\omega} 
({\bf{\omega}} - \tilde{W} {\bf{c}})
\end{equation}
It is straightforward to show 
that the value of the parameters that minimize $\chi^2$ are given by
\begin{equation}
{\bf{c}} =  \Omega^{-1}  \tilde{W}^T \tilde{C}_\omega^{-1} \omega
\end{equation}
where $\Omega = \tilde{W}^T \tilde{C}_\omega^{-1} \tilde{W}.$
Because of the large attenuation of
the ACF at large angles, we chose to 
fit to only those data points with $\theta_i \le 9^{\circ}$, i.e., $i \le 8$, 
even though there appears to be statistically significant structure out to 
$\theta \sim 13^{\circ}$. 
The results of the fit of the $\epsilon = 1$ model to the ACF of the residuals of 
Map A are listed in Table \ref{tab:acf} and plotted in Figure \ref{fig:acf}.

It is also straightforward to show that the correlation matrix
of the fit parameters is given by, $ \langle \delta c_n \delta c_m \rangle
= \Omega_{nm}^{-1}$
so the errors given in Table \ref{tab:acf}
 are given by $\sigma_{c_n}^2 = \Omega_{nn}^{-1}$
and the normalized correlation coefficients by 
$r_{nm} = \Omega_{nm}^{-1}/(\Omega_{nn}^{-1}\Omega_{mm}^{-1})^{1/2}$.  
The correlations are as
expected, i.e., $\omega_{intr}$ and $\omega_{psf}$ are highly correlated,  
while $\omega_s$ is relatively
uncorrelated with the other two parameters.

From the results in Table \ref{tab:acf} 
it appears that intrinsic correlations in the X-ray
background are detected at the 4 $\sigma$ level for sources with flux levels 
below
$3 \times 10^{-11} erg~s^{-1} cm^{-2}$.  Of course, if we have not successfully
eliminated sources with fluxes larger than this, then the clustering amplitude
might well be artificially inflated.  It was for this reason that we used the
RASS catalog to identify and remove additional sources with intensities
$\gs 3 \times 10^{-11} erg~s^{-1} cm^{-2}$, the result of which was Map B
(see \S2). 
The fits to Map B are also listed in
Table \ref{tab:acf}.
The clustering amplitude, $\omega_{intr}$, of the fits to this modified map
were only $11\%$ less than that of Map A, i.e., considerably less than 1 
$\sigma$. The $\chi^2$s of both fits are acceptable.

These results are not very sensitive to the attenuation corrections.
If they are removed from the model, 
the resulting amplitude of the fit clustering 
coefficient only decreases by $\sim 20\%$, less than $1~\sigma$.
It should be noted that the corrections were
relatively small (an average attenuation factor of 0.83 ranging from 1.0 at
$\theta = 0^{\circ}$ to 0.55 for $\theta = 9^{\circ}$)
and in all cases, the attenuation was less than the error bar 
of the corresponding data point.  If data points 
with $\theta > 9^{\circ}$ are included, the fits become 
more sensitive to the attenuation corrections which are, in turn,  
quite sensitive to the 8-parameter fit.

\begin{table*}[ht]
\begin{center}
\begin{tabular}{|c|r|r|r|r|r|r|r|}
\multicolumn{1}{l}{Map} & \multicolumn{1}{c}{$\omega_s \times 10^4$} &
\multicolumn{1}{c}{$\omega_{PSF} \times 10^4$} &
\multicolumn{1}{c}{$\omega_{intr} \times 10^4$} & 
\multicolumn{1}{c}{$\chi_5^2$} & \multicolumn{1}{c}{$r_{12}$} & 
\multicolumn{1}{c}{$r_{13}$} & \multicolumn{1}{c}{$r_{23}$} \\
\cline{1-8}  
A & 7.71 $\pm$ 0.15 & 8.63 $\pm$ 0.59 & 2.64 $\pm$ 0.65 & 2.3/5 & -0.19 & -0.04 & -0.83 \\
\cline{1-8}  
B & 7.57 $\pm$ 0.15 & 8.22 $\pm$ 0.59 & 2.33 $\pm$ 0.65 & 3.6/5 & -0.19 & -0.04 & -0.83 \\
\cline{1-8}
\end{tabular}
\end{center}
\caption{}
\label{tab:acf} 
{Fit model parameters for Map A with 55\% sky coverage and Map B
with 52\% sky coverage after removing additional ROSAT sources (see \S2).
The intrinsic fluctuations are modelled as $\omega \propto \theta^{-1}$.}
\end{table*}

Figure \ref{fig:resid} is a plot of the residuals of the fit to the ACF of Map A
for $\theta \le 9^{\circ}$ and of the uncorrected ACF from $10^{\circ}$ to $180^{\circ}$.
The vertical scale is the same as for Figure \ref{fig:acf}.  
The \emph{rms} of these 140 data points 
is $1 \times 10^{-5}$ and it is clear that there is very little residual
structure at levels exceeding this value.  
The observed correlation function for $\theta > 10^{\circ}$
in Figure \ref{fig:resid} is entirely consistent with the noise levels 
determined from the Monte Carlo simulations: the \emph{rms} of $\omega/\sigma$ 
is 1.03, indicating that there is 
no evidence for intrinsic fluctuations on these scales. 
We also take this as an indication that the 
errors are reasonably well characterized by the Monte Carlo calculation.  The 
variance of the photon shot noise, ${\sigma_{s}}^2 = \omega_{s} {\bar{I}}^2$, 
is consistent with that expected from 
photon counting statistics only (Jahoda 2001).  It should be noted that not
only the shape but the amplitude of the beam smearing contribution, $\omega_{PSF}$, 
can be computed from source counts as a function of $2-10~keV$ flux.  We will 
argue in \S6 that, while our fitted value is consistent with current number 
counts, the latter are not yet accurate enough to correct the data.

Figure \ref{fig:intrin} shows the model of the intrinsic clustering, 
$\omega_{intr}$, compared to 
the data with both the shot noise and PSF
component removed.  
The model curve is not plotted beyond $\theta = 9^{\circ}$ since the attenuation
factor due to the 8-parameter fit corrections are large and uncertain at larger
angles.
The amplitude of $\omega_{intr}$ is sensitive to the 
exponent in the assumed power law for the intrinsic correlations.  For example,
the fit amplitude for a  $\theta^{-1.6}$ power law the amplitude is a factor of
$\sim 2$ larger than for a $\theta^{-0.8}$ power law.  However, for a range of fits
with $0.8 \le \epsilon \le 1.6$, the values of $\omega_{intr} (\theta)$ at
$\theta = 4.5^{\circ}$ are all within $\pm 3\%$ of each other.  Therefore, we chose
to normalize the X-ray ACF at $4.5^{\circ}$ when comparing to cosmological models
(see \S7.1).  The $\chi^2s$ are reasonable
for all these fits.  

The bottom line is that there is fairly strong 
evidence for intrinsic clustering on these angular scales at the level of
$\omega_{intr} \sim 3.6 \times 10^{-4} \theta^{-1}$ (see \S6.1).  While the exponent
is $\epsilon \sim 1$, it is not strongly constrained.
The implications of the intrinsic clustering of the X-ray background will be
discussed in \S7.

\section{Comparisons with Previous work}

\subsection{Clustering}

As mentioned above, the component of the ACF due to beam smearing, 
$\omega_{PSF}$, can be determined with no free parameters if the 
flux limited number counts of X-ray sources are known.  While such 
counts are still relatively inaccurate for our purposes,  we did 
check to see if our results are consistent with current data.   Over a 
restricted flux range, the X-ray number counts, $N(<S)$, are reasonably approximated 
by a power law, i.e., $N(< S) = K~S^{-\gamma}$ where $S$ is the flux of the source.
It is straightforward to show that the variance of flux due to a Poisson distribution 
of sources is
\begin{equation}
{\sigma_{PSF}^2} (0) = \pi {\sigma_p}^2~A^2~\gamma~K~S^{2-\gamma}/(2-\gamma)
\end{equation}
where $S$ is the upper limit of source flux, $\sigma_p$ is the Gaussian width 
of the PSF and $A^{-1}$ is the flux of a point
source that results in a peak signal of $1~TOT~count~s^{-1}$ in our composite 
map.  In our case, $A = 2.05 \times 10^{10}erg^{-1}~s~cm^2$.  Using the 
$BeppoSAX~2-10~kev$ number count data of Giommi, Perri, \& Fiore (2000), the 
$Chandra$ data of Mushotzky et al. (2000), and the $HEAO1~A2$ data of Piccinotti
et al. (1982), we constructed a piecewise power-law $N(< S)$ for the range 
$6 \times 10^{-16} < S < 3 \times 10^{-11} erg~s^{-1}cm^{-2}$ and computed
$\omega_{PSF}(0) = \sigma^2_{PSF}/\bar{I}^2$ to be $8.4 \times 10^{-4}$.  
The close agreement of this 
value with those in Table \ref{tab:acf} 
is fortuitous 
given that the value depends most 
sensitively on the number counts at large fluxes which are the most unreliable,
typically accurate only to within a factor of two. 
However, it is clear that the $\omega_{PSF}$ of 
Table \ref{tab:acf} are quite 
consistent with the existing number count data.

The results of \S5 clearly indicate the 
presence of intrinsic clustering in the 
XRB.  Mindful that the detection is only $\sim 4~\sigma$, we tentatively assume 
that the ACF has an amplitude of  $\omega_{intr} \sim 2.5 \times 10^{-4}$ (the
average of the two values in Table \ref{tab:acf})
and is consistent with a $\theta^{-1}$ functional dependence.  It is straight 
forward to relate this to the underlying correlation amplitude, $\theta_0$. 
The variance of correlations smoothed by a beam of Gaussian size $\sigma$ is
given by 
\begin{equation}
\omega_{intr}(0) = {\Gamma(1 - \epsilon/2) \over 2^{\epsilon}}
\left({\theta_0 \over \sigma} \right)^\epsilon.  
\end{equation}
When $\epsilon = 1$, then $\theta_0 = 2 \sigma \omega_{intr}/\pi^{1/2}$.  
Using this, we find 
\begin{equation}
\omega_{XRB}(\theta) \simeq 3.6 \times 10^{-4}~\theta^{-1}
\label{eqn:acf}
\end{equation}
where $\theta$ is measured in degrees and the normalization is such that 
$\omega_{XRB}(0) = \langle {\delta I}^2 \rangle / \bar{I}^2$.

For comparison, this amplitude is about a factor of three below the $2~\sigma$
upper limit determined by Carrera et al. (1993) obtained with Ginga data 
($4-12~keV$) for angular scales between 0.2 and 2.0 degrees.  
The detection of a significant correlation in the $HEAO1~A2$ data at the level
of $3 \times 10^{-5}$ at $\theta = 10^{\circ}$ by Mushotzky and Jahoda (1992)
was later attributed to structure near the super-Galactic plane (Jahoda 1993).
To check for this effect in the present analysis, we masked all pixels within
15 and 20 degrees from the super-Galactic plane.  The results were 
indistinguishable from those of Table \ref{tab:acf}.  It is
interesting that the clustering indicated in Eq. (6-3) 
is consistent with this level of fluctuations at $10^{\circ}$;
however, our sensitivity has begun to decline 
significantly at $10^{\circ}$ due to the fit for large scale structures.
A correlation analysis of ROSAT soft X-ray background by Soltan 
et al. (1996) detected correlations about an order of magnitude larger than
indicated in Eq. (6-3).  
Even considering that the ROSAT band ($0.5-2.0~keV$) is
distinct from the HEAO band, it is difficult to imagine that the two 
correlation functions could be so disparate unless the lower energy analysis 
is contaminated by the Galaxy.

While there has yet to be a definitive detection of the clustering of hard 
X-ray sources, a recent deep Chandra survey of 159 sources shows a positive 
correlation of source number counts on angular scales of 5 to 100 arcsec 
(Giacconi et al. 2001).  
Although the signal to noise is low and dependent on source flux, 
the implied number count ACF is roughly consistent with
$\omega_N (\theta) \sim 3 \times 10^{-3}\theta^{-1}$ where 
$\omega_N (0) \equiv \langle {\delta N}^2 \rangle / \bar{N}^2$ and $\bar{N}$ is
the mean surface density of sources.  This is consistent with the 
correlation function determined by Vikhlinin \& Forman (1995) for sources
identified within ROSAT PSPC deep pointings.  A direct comparison between
the Chandra result with that of Eq. (6-3) is complicated by the
more than one hundred times difference 
in the angular scales of the two analyses.
It is doubtful that a single powerlaw model is adequate over this range.
Furthermore, one is a luminosity ACF while the other is a flux limited,
number count ACF.  Relating the two requires understanding 
the luminosity function and its evolution 
as well as how the X-ray bias depends on scale.
For these reasons, a direct comparison would
be difficult to interpret.  We only note in passing that the small angular
scale ACF is a factor of eight larger than that of Eq. (6-3)
assuming a $\theta^{-1}$ dependence.

Finally, the recent harmonic analysis of the $HEAO1~A2$ data by Scharf et al.
(2000) yielded a positive detection of structure in the XRB out to harmonic
order $l \sim 10$.  The present analysis looks at similar maps, so the results 
should be comparable. 
A direct comparison 
is complicated by the differences in analysis techniques, masking and 
corrections to the map. 
A rough comparison can be made by performing a Legendre transform on the 
$\theta^{-1}$ ACF model of Eq. (6-3).
The ACF can be expressed in terms of Legendre polynomials as
\begin{equation}
\omega(\theta) = {{1}\over{4\pi}} \sum_\ell (2\ell+1){C_\ell} P_\ell(\cos\theta)
\end{equation}
where the ${C_l}$ constitute the angular power spectrum.  Taking the Legendre
transform
\begin{equation}
{C_\ell} = 2 \pi \int_{-1}^1 \omega(\theta)~P_\ell(\cos \theta)~d(\cos\theta) 
\end{equation}
where $P_\ell(\theta)$ is the Legendre polynomial of order $\ell$.  
Substituting the 
$\theta^{-1}$ model into this expression results in power spectrum 
coefficients, $C_\ell \simeq 4\times 10^{-5}/\ell$ for $\ell \sim 5.$
Note that for $\ell \sim 5$, this expression is relatively insensitive 
to the index $\epsilon$ in the expression for $\omega_{intr}$.
While these values are highly uncertain,
they are comparable to those found by Scharf et al. (2000) when the sky 
coverage and differences 
in notation are accounted for. 
Considering 
the low signal to noise of the data as well as the differences in the two 
analyses, a more detailed comparison would not be particularly useful.

\subsection{The Dipole}

Scharf et al. (2000) also searched for the XRB dipole using a 
$HEAO1~A2$ map and similar methods to those described in \S4.
They claim a detection of an intrinsic dipole with
amplitude $\Delta  \sim 0.0065$, though with a rather large region of
uncertainty, i.e., $0.0023 \ls \Delta \ls 0.0085$, and in a 
direction about $80^{\circ}$ from that of the Compton-Getting dipole, in the 
general direction of the Galactic center.  
They used  
the $3^\circ \times 1.5^\circ$ $HEAO1~A2$ map restricted 
to regions further
than $22^{\circ}$ from the Galactic plane.  In addition, regions about sources
with fluxes greater than $2 \times 10^{-11} erg~s^{-1} cm^{-2}$ were cut from 
the map.  The sky coverage and the level of source removal closely correspond 
to those of our Map C.   
However, since we used a combination of 
the $3^\circ \times 1.5^\circ$ and $3^\circ \times 3^\circ$ maps, our map has 
significantly less ($\sim 1/\sqrt 3$) photon shot noise.
While our analyses are similar, there are some significant differences: 
they corrected the map beforehand for linear instrument drift and Galaxy 
emission while we fit for those 
components simultaneously with the dipole and with emission from the local 
supercluster which they ignore.  

The upper limit on the intrinsic dipole we find is about the 
same amplitude as
the Compton-Getting dipole, i.e., $\Delta < 0.0052$ at the 95 \% C.L.  
Thus we exclude roughly the upper half of the Scharf et al. range and believe   
that their claim of a 
detection is probably an overstatement.  While Scharf et al. do not take
into account emission from the plane of the local supercluster, even if 
we leave that component out of our fit, the dipole moment (including the
C-G dipole) increases by only $0.005~TOT~cts~s^{-1}$ and is still consistent
with the C-G dipole.  The upper limit on the intrinsic dipole with this
fit is determined primarily by the noise in the fit and is not significantly
different from that value given above.  
It is difficult to understand how their quoted errors could be two to four
times 
less than those quoted in Table \ref{tab:dipole}.  The shot noise variance in 
the map they used was three times greater than 
in our combination map and so one 
would expect their errors would be somewhat larger than those above. 
They performed only 
a four parameter fit (offset plus dipole) which would result in a slight 
reduction of error; however, this is a bit misleading since their Galaxy model
and linear time drift are derived from essentially the same data set.  
It is possible that their lower errors could 
result from ignoring correlations in the noise (our detected ACF). 
In any
case, we find no evidence for an intrinsic dipole moment in the XRB.

\section{Implications for Cosmology}

\subsection{Clustering and Bias in the X-ray Background}

The observed X-ray auto-correlation can be compared to the matter 
auto-correlation predicted by a given cosmological model. 
The linear bias factor for the
X-rays can then be determined
by normalizing to the observed CMB anisotropies.
Since X-rays arise at such high redshifts,
the fluctuations we measure are on scales $\lambda \sim 100 h^{-1} Mpc$,
comparable to those constrained by the CMB, i.e.,
on wavelengths that entered the horizon about the time of matter domination.

The predicted X-ray ACF depends on both the cosmological model and on the
model for how the X-ray sources are distributed in redshift,
which is constrained by observed number counts and the redshift
measurements of discrete sources.  We use the redshift
distribution described in
Boughn, Crittenden and Turok (1998), based on the unified AGN model of
Comastri et al. (1995).  (See also the more recent analysis by
Gilli et al. (2001).)
While we will not reproduce those calculations here,
the basic result is that the XRB intensity is
thought to arise fairly uniformly in redshift out to $z=4$.
Our results here are not very sensitive to the precise details of this
distribution.

Another issue in the calculation of the power spectrum is the possible
time dependence of the linear bias.  Some recent studies indicate that the
bias is tied to the growth of fluctuations and may have
been higher at large redshift (Fry 1996, Tegmark \& Peebles 1998).  
For the purposes of the power spectrum, an evolving bias 
will have the same effect as changing the source redshift distribution.
Again, our results are not strongly dependent on these uncertainties, but
they comprise an important challenge to using the X-ray fluctuation studies
to make precision tests of cosmology.

\begin{figure}
\centerline{\includegraphics*[width=3.0in]{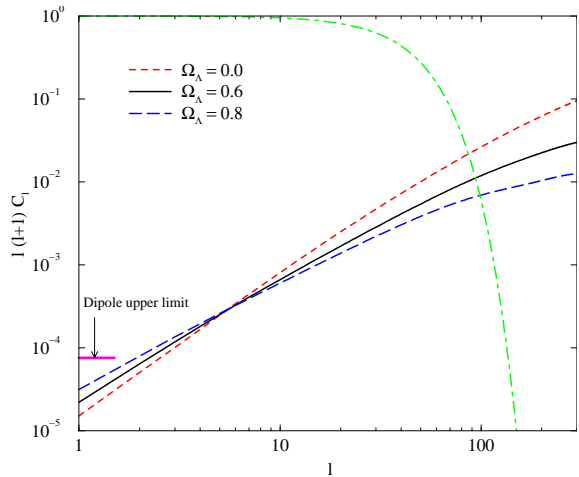}}
\caption{The power spectrum for a range of cosmologies normalized
to the observations ($H_0 = 70~km~s^{-1}Mpc^{-1}$.)  
The various cosmologies show a 
range of slopes, from $1.1 < \epsilon < 1.6 $ and the observations fix them 
at $\ell \simeq 5$.  Also shown is the 95 \% upper limit from the dipole, 
excluding cosmic variance.  With cosmic variance, the limit shown is at 
the 80 \% confidence level, and the 95 \% upper limit is four times higher. 
The green line shows the suppression arising from beam smoothing, which 
smoothes scales $\ell > 50$. 
}
\label{fig:x-cls}
\end{figure}

Figure \ref{fig:x-cls} shows the predicted XRB power spectrum, normalized
to our observations.   On the scales of interest, the predicted spectra are
fairly featureless, and reasonably described by a power law in $\ell$,
$C_\ell \propto \ell^{\epsilon -2}$, which corresponds to the correlation
of the form $\omega(\theta) = (\theta_0/\theta)^\epsilon$.
For the models of interest, $1.1 < \epsilon < 1.6$ for $\ell < 100$, and
decreases at higher $\ell$ (smaller separations).
Note that the spectra calculated by Treyer et al. (1998) appear to
be consistent with our
findings, suggesting $\epsilon = 1.2$.
The precise index $\epsilon$ depends on the position of the power spectrum peak 
which is determined by the shape parameter, 
$\Gamma \simeq \Omega_m h.$  Larger values of 
$\Gamma$ imply more small scale power and thus higher $\epsilon.$ 

For simplicity, we normalize to the X-ray correlation function at $4.5^{\circ}$,
$\omega(4.5^\circ)= 1.0 \pm 0.25\times 10^{-4}$.
This separation is large enough to be independent of 
the PSF contribution to the ACF, but 
not so large that the attenuation from the large scale fits 
becomes significant. 
Also, the value of the fit ACF at $4.5^{\circ}$ is nearly independent of the 
index $\epsilon$ (see \S5).
As can be seen from Figure \ref{fig:x-cls}, this normalization fixes
the power spectrum at $\ell \simeq 5-7$.

We normalize the fluctuations to the COBE power spectrum as determined
by Bond, Jaffe \& Knox (1998).  
However, it should be noted that fits to smaller angular CMB fluctuations
indicate that using COBE alone
may somewhat overestimate the matter fluctuation level (Lahav et al. 2002). 
The biases derived from the models appear to be largely insensitive to 
the matter density.  This is due to a cancellation of two effects: the CMB 
normalization and the power spectrum shape (White \& Bunn 1995). 
The biases are roughly inversely 
proportional to $h$.  Typical biases appear 
to be $b_X = 2.3 \pm 0.3 (0.7/h)^{0.9}$, 
increasing slightly as $\Gamma$ decreases and the 
peak of the power spectrum moves to larger scales.

\subsection{The Intrinsic X-ray Dipole } 

The theoretical models normalized to our observations 
predict the intrinsic power on a wide range of scales, 
assuming the X-ray bias is scale independent.
In particular, these models give a prediction for the variance of 
the intrinsic dipole moment.  We can compare our model predictions to the 
the upper limit for the intrinsic dipole to see if we should have observed it 
in the X-ray map. 

The dipole amplitude in the $\hat{z}$ direction is related to the spherical 
harmonic amplitude by $\Delta_z = \sqrt{3/4\pi} a_{10}$.  Thus, the expected dipole 
amplitude is related to the power spectrum by 
\begin{equation} 
\langle \Delta^2 \rangle = 
3 \times {3 \over 4 \pi} \langle |a_{1m}|^2 \rangle = {9 \over 4 \pi} C_1.  
\end{equation} 
Note that there is considerable cosmic variance on this, as it is 
estimated with only three independent numbers; $\delta C_1/ C_1 = (2/3)^{1/2}$,
which corresponds to a 40 \% uncertainty in the amplitude of the dipole.  

Also shown in Figure \ref{fig:x-cls} is the level of our dipole limit, 
translated using equation (7-1),   
which corresponds to $C_1 < 3.8 \times 10^{-5}$. 
While our limit on the dipole limit is 
at the 95 \% C.L., this translates to a 80 \% C. L. limit on the variance 
$C_1$ when cosmic variance is included.    
As discussed above, the 95 \% upper limit is four times weaker when cosmic 
variance is included, $C_1 < 1.5 \times 10^{-4}$.   
Normalized to the ACF, all the theories are easily 
compatible with the $C_1$ bound.  
The large 
cosmic variance associated with the dipole makes it difficult to rule out any 
cosmological models.  

With our detected level of clustering, typical theories would predict 
a dipole amplitude of $\Delta \simeq 0.003$. 
While the theories are not in conflict 
with the dipole range claimed by Scharf et al., they strongly prefer the lower 
end of their range, even for the most shallow of the models.  
A dipole amplitude $\Delta \ge 0.005$ would be very unlikely from the models,
indicating either a significantly higher bias than we find or a model 
with more large scale power ($\epsilon \le 1.1$).  

\subsection{The dipole and bulk motions} 

The dipole of the X-ray 
background provides another independent test of the large scale X-ray bias
through its relation to our peculiar velocity (e.g. see Scharf et al. (2000)
and references therein.)
Like the gravitational force, 
the flux of a nearby source drops off as an inverse-square law, so  
the dipole in the X-ray flux is proportional to the X-ray bias times the 
gravitational force produced by nearby matter. 
Our peculiar motion is a result of this force, 
and is related to the gravitational 
acceleration by a factor which depends on the matter density.  

In typical CDM cosmologies, the dipole and our peculiar velocity 
arise due to matter at fairly low redshifts ($z < 0.1$).
If this is the case, it is straight forward to relate their amplitudes. 
Following the notation of Scharf et al., we define 
$D^\alpha = \int d\Omega I(\hat{\bf{n}}) \hat{n}^\alpha = 
4\pi \bar{I} \Delta^\alpha/3$.  Using linear perturbation theory, 
one can show the local bulk flow is  
\begin{equation} 
v^\alpha = {{H_0 f} \over {b_X(0) \rho_X (0)}} D^\alpha,  
\end{equation}  
where $\rho_X (0)$ is the local X-ray luminosity density, $b_X(0)$ is the 
local X-ray bias and $f \simeq \Omega_m^{0.6}$ is related to the growth of 
linear perturbations (Peebles 1993).  
From the mean observed intensity (Gendreau et al. 1995) and the local X-ray
luminosity density (Miyaji et al. 1994) we find that
$\bar{I} \simeq 2.4 \rho_X (0) c/4 \pi H_0$.  
This implies that 
\begin{equation} 
|v| \simeq 2.4 \times 10^5 \Delta \, {\Omega_m^{0.6} \over b_X(0)} 
\, \rm{km\,s^{-1}}. 
\end{equation}
This relation was derived by Scharf et al. (2000), though their numerical factor
was computed from a fiducial model rather than directly from
the observations, as above.  In any case, the uncertainty in  $\rho_X (0)$
is considerable, $6 \times 10^{38} erg\, s^{-1}Mpc^{-3} < \rho_X (0)
< 15 \times 10^{38}erg\, s^{-1}Mpc^{-3}$ (Miyaji et al. 1994), and
so is the uncertainty in this relation.

Our maps have bright sources removed, which correspond to 
nearby sources out to $60 h^{-1} Mpc$.  Thus, we need to compare our 
dipole limit to the motion 
of a sphere of this radius centered on us.  
Typical velocity measurements on this scale find 
a bulk velocity of $v_{60} \simeq 300 \pm 100$ km/s 
(see Scharf et al. (2000) for a summary.) 
With our dipole limit, this implies that 
$\Omega_m^{0.6}/b_X(0) \gs 0.24 \pm 0.08$ where the uncertainty in 
$\bar{I}/\rho_X (0)$ is not included.
This constraint is independent 
of cosmic variance issues. 
While the diameter of the local (Virgo) supercluster is generally considered
to be in on the order of $40 h^{-1}$ to $50 h^{-1}~Mpc$ (e.g., Davis et al. 1980),
there is evidence that the overdensity in the Supergalactic plane extends
significantly beyond $60h^{-1}~Mpc$ (Lahav et al. 2000).  One might, therefore,
suspect that our correction for emission from the local supercluster might
effectively remove sources at distances greater than $60h^{-1}~Mpc$ in that plane.
In any case, that correction made very little difference in dipole fits (see
\S 6.2) so our conclusions remain the same.

Note that this limit could potentially conflict with previous determinations 
by Miyaji (1994) who found $\Omega_m^{0.6}/b_X(0) = f_{45}/3.5,$ where 
$f_{45}$ is the fraction of 
gravitational acceleration arising from $R \le 45 h^{-1}$ Mpc. 
This is consistent only for $f_{45} \sim 1$, which is larger than is 
usually assumed ($f_{45} \sim 0.5$).  However, this limit comes from studies of
a fairly small sample (16) of X-ray selected AGN and is subject to significant 
uncertainties of its own. 

For typical biases suggested by the observed clustering ($b_X \sim 2.3$), our
constraint suggests a somewhat high matter density, $\Omega_m > 0.37$, 
for $v_{60} \simeq 300$ km/s.   
This is consistent with the ISW constraint discussed below and also with 
previous analyses of bulk velocities which tend to 
indicate higher $\Omega_m$. 
However, if the bulk velocity is smaller and/or $\rho_X (0)$ larger, 
this constraint is weakened.  
In addition, we have assumed a constant X-ray bias.  If the bias evolves with 
redshift, then the local value could be considerably smaller which would also 
weaken this bound.

\subsection{The Integrated Sachs-Wolfe Effect and 
$\Omega_{\Lambda}$}

In models where the matter density is less than unity, 
microwave background fluctuations 
can be created very recently by the evolution of the 
linear gravitational potential.  This is known as the late time 
integrated Sachs-Wolfe (ISW) effect.  Photons gain energy as they fall into 
a potential well, and loose a similar amount of energy as they exit. 
However, if the potential evolves significantly as the photon passes through, 
the energy of the photons will be changed, leaving an imprint on the 
CMB sky.   The spectrum is modified most on large scales where 
the photons receive the largest changes. 

The CMB anisotropies created in this way are naturally correlated with 
the gravitational potential.  
Thus, we expect to see correlations between the CMB and tracers of the 
local ($z \sim 2$) gravitational potential such as the X-ray background
(Crittenden \& Turok 1996).  
These correlations are primarily on large scales such as those probed 
by the HEAO survey.  

In an earlier paper, we searched for a correlation between the HEAO maps and 
maps of the CMB sky produced by COBE.   We failed to find such a cross 
correlation and were able to use our limit to constrain the matter density 
and the X-ray bias (Boughn, Crittenden \& Turok 1998, hereafter BCT).  
However, translating our measurement into a cosmological bound was 
ambiguous because the level of the intrinsic structure of the XRB 
was unknown at the time.  
With the observation of the X-ray ACF presented here, we are in a position 
to revisit the cosmological limits implied by these measurements.  

To make cosmological constraints, we compare the observed X-ray/CMB
cross correlation
to those predicted by $\Lambda CDM$ models. 
As above, we normalize the CMB fluctuations using the band powers of COBE 
(Bond, Jaffe \& Knox 1998) and also normalize the X-ray fluctuations 
as discussed in \S7.1.   
The cross correlation analysis of BCT was performed with a coarser 
pixelization ($2.6^\circ \times 2.6^\circ$) than the ACF discussed above. 
We include this effect by using the numerically calculated pixelization 
window function.  The COBE PSF used was that found by 
Kneissl \& Smoot (1993) and we used a $2.9^\circ$ FWHM Gaussian for the 
underlying X-ray PSF (recall that the $3.04^\circ$ FWHM beam found 
above includes a $1.3^\circ \times 1.3^\circ$ pixelization.) 

\begin{figure}
\centerline{\includegraphics*[width=3.0in]{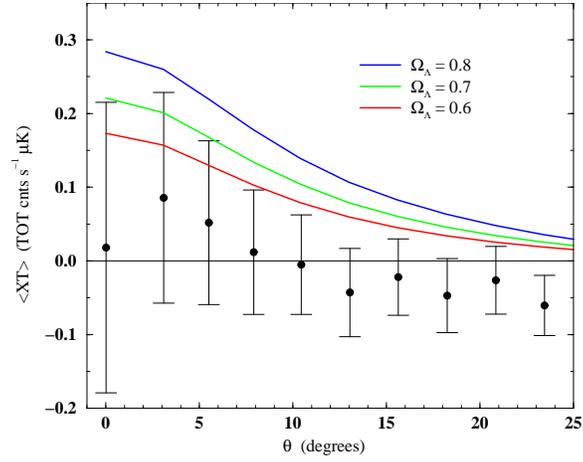}}
\caption{ The calculated X-ray/CMB cross correlation.  
The error bars are highly correlated. Also shown 
are the predictions for three $\Lambda-CDM$ models with varying $\Omega_\Lambda$
($H_0 = 70~km~s^{-1}Mpc^{-1}$.)
} 
\label{fig:cross} 
\end{figure}

The calculation of the HEAO-COBE cross correlation was discussed in BCT and 
has not changed.  The results are shown in Figure \ref{fig:cross}, along with 
predictions for three different values of $\Omega_\Lambda$.  
While the X-ray bias depends strongly on the Hubble parameter, the 
predicted cross correlation is only weakly dependent on it, changing only 
10\% for reasonable values of $H_0$.  The cross correlation depends primarily 
on $\Omega_\Lambda$; no correlation is expected if there is no cosmological 
constant and the ISW effect increases as $\Omega_\Lambda$ grows. 
The error bars in  Figure \ref{fig:cross} are calculated from Monte Carlo 
simulations and arise primarily due to cosmic variance in the observed 
correlation.  The error bars are significantly correlated.

\begin{figure}
\centerline{\includegraphics*[width=3.0in]{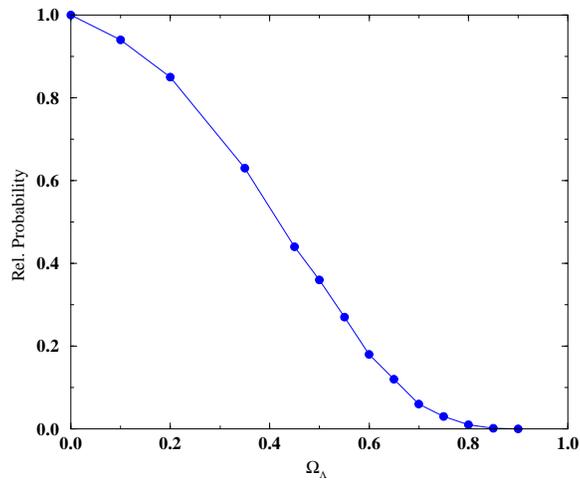}}
\caption{  The relative probability of the observed cross correlation
for varying cosmological constant, with the Hubble constant fixed  
($H_0 = 70~km~s^{-1}Mpc^{-1}$.)  The best fit is for no correlation. 
} 
\label{fig:rprob} 
\end{figure}

The observed correlation is most consistent with there being no intrinsic 
cross correlation ($\Omega_\Lambda =0.0$).   
We set limits by calculating the likelihood of a model relative to this 
no correlation model.  Using the frequentist criterion used in BCT, 
$\Omega_\Lambda \le 0.65$ at the 98\% C.L., $\Omega_\Lambda \le 0.60$ 
at the 95\%. C.L.  Almost identical limits arise from a Bayesian approach, where
the relative likelihoods are marginalized over, assuming a constant 
prior for $\Omega_\Lambda \ge 0$.  Figure \ref{fig:rprob}
shows a one-dimensional slice through the likelihood surface, where only the
cross correlation information has been used to calculate the likelihood. 

One of the major assumptions we made in interpreting the above result is how 
the sources of the XRB are distributed in redshift.  It is likely that current
models of the luminosity function will have to be substantially modified as
further deep observations of the sources of the XRB are made.  However, as pointed
out above, the ISW is relatively insensitive to the exact shape of the redshift 
distribution of luminosity.  If the true distribution includes a substantial
fraction of the luminosity at redshifts greater than 1, then the above results 
will not change dramatically.  On the other hand, 
our constraint on $\Omega_\Lambda$
is quite sensitive to the value of the bias parameter.  If the sources of the XRB
should turn out to be unbiased, i.e., $b_X = 1$, then the constraint on 
$\Omega_\Lambda$ could be weakened dramatically. We hasten to add that such a low
bias would require that the ACF of Figure 5 be reduced by more than a factor of
four, which seems unlikely.  Previous determinations of 
X-ray bias have resulted in a wide range of values, $1 < b_X < 7$ (see Barcons
et al. 2000 and references therein).  Its clear that firming up the value of
$b_X$ and determining how it varies with scale and redshift will be required before
the ISW effect can be unambiguously interpreted. 

The above limit may be compared to what we found from cross correlating 
COBE with the NVSS radio galaxy survey (Boughn \& Crittenden 2002).  
There we also found no evidence
for correlations, and were able to put a 95 \% C.L. limit of 
$\Omega_\Lambda \le 0.74$, with some weak dependence on the Hubble constant. 
While the above limit provides important confirmation of that result, it should 
be noted that these two limits are not entirely independent.  
Radio galaxies and the X-ray background are, indeed, correlated 
with each other (Boughn 1998).

An important source of noise in the cross correlation of Figure 7 is instrument
noise of the COBE DMR receivers.  In addition, the relatively poor angular 
resolution of the COBE radiometers reduce, somewhat, the amplitude of the ISW
signal.  Therefore, some improvement can be expected by repeating the analysis
on future CMB maps, such as that soon to be produced by NASA's MAP satellite
mission.  If such an analysis still finds the absence of an ISW effect, then
the current $\Lambda CDM$ model would be in serious conflict with observational
data if the X-ray bias can be similarly constrained.  
On the other hand, a positive detection would provide important evidence 
about the dynamics of the universe even if the X-ray bias remains uncertain.

\section{Conclusions}

By carefully reconstructing the HEAO beam and analysing its auto-correlation
function, we have been able to 
confirm the presence of intrinsic clustering in the X-ray background. 
This gives independent verification of the multipole analysis of 
Scharf et al. (2000) and the level of clustering we see is comparable. 
The clustering we see is in excess of that predicted by standard cold 
dark matter models and indicates that some biasing is needed.  
The amount of biasing required depends on the cosmological 
model and on how the bias
evolves over time; if the bias is constant, typical models indicate that 
$b_X \simeq 2.$  The biases of galaxies, clusters of galaxies, radio sources,
and quasars have yet to be adequately characterized and so whether or not the 
above X-ray bias is excessive is a question that, for the present, remains 
unanswered.

We have also confirmed, at the 2-3 $\sigma$ level,  
the detection of the Compton-Getting dipole in the X-ray background due 
to the Earth's motion with respect to the rest frame of the CMB.  
However, we have been unable to confirm the presence of an intrinsic dipole
in the XRB and have actually been able to exclude a significant part of 
the range reported by Scharf et al. (2000).  While our dipole 
limit is still too small to conflict with any of the favored CDM models, 
combining our  
dipole limit with observations of the local bulk flow enable us to constrain 
$\Omega_m^{0.6}/b_X(0) > 0.24$.  For constant bias models, this suggests 
a relatively large matter density, as is also seen in for other velocity studies;
however, the uncertainty in this limit is still considerable. 

With the observed X-ray clustering, large $\Lambda-CDM$ models 
predict a detectable correlation with the cosmic microwave background 
arising via the integrated Sachs-Wolfe effect.  
That we have not observed this effect suggests $\Omega_\Lambda \ls 0.60$. 
This is beginning to conflict with models preferred 
by a combination of CMB, LSS and 
SNIA data (e.g., de Bernardis et al. 2000 \& Bahcall et al. 1999).
 
This work gives strong motivation for further 
observations of the large scale structure of the hard X-ray background.  
Better measurements of the full sky 
XRB anisotropy are needed, as is more information 
about the redshift distribution of the X-ray sources. 
This will be essential for cross correlation with the new CMB data from the 
MAP satellite and to bridge the gap between the CMB scales and those 
probed by galaxy surveys such as 2-dF and SDSS.

\begin{acknowledgments}
We would like to acknowledge Keith Jahoda who is responsible for constructing
the HEAO1 A2 X-ray map and who provided us with several data-handling 
programs.  We also thank Neil Turok for useful discussions, Ed Groth for a 
variety of analysis programs, and Steve Raible for his help with some of the
analysis programs.  RC acknowledges support from a PPARC Advanced
Fellowship.  This work was supported in part by NASA grant NAG5-9285.
\end{acknowledgments}

\newpage 
\newpage

\end{document}